\documentclass[]{aa}
\usepackage{graphicx}
\usepackage[below]{placeins}

\begin{document}

\titlerunning{The oxygen-K absorption edge}
\title{The interstellar oxygen-K absorption edge as 
       observed by XMM-Newton}

\subtitle{Separation of instrumental and interstellar components}

\author{C.P. de Vries \inst{1} \and
	J.W. den Herder \inst{1} \and
	J.S. Kaastra \inst{1} \and
	F.B. Paerels \inst{2} \and
	A.J. den Boggende \inst{1} \and
	A.P. Rasmussen \inst{2}
	}

\offprints{C.P. de Vries, \\
	   \email{C.P.de.Vries@sron.nl}}

\institute{SRON, National Institute for Space Research, 
		Sorbonnelaan 2, 3584 CA Utrecht, The Netherlands
	   \and
	   Columbia Astrophysics Laboratory, Columbia University,
	   550 West $\mathrm{120^{th}}$ Street, New York, NY 10027, USA
	   }

\date{Received $\prec date1 \succ$, accepted  $\prec date2 \succ$}

\abstract{
High resolution X-ray spectra of the Reflection Grating Spectrometer
(RGS) on board the XMM satellite are used to resolve the oxygen K
absorption edge. By combining spectra of low and high extinction
sources, the observed absorption edge can be split in the true
interstellar (ISM) extinction and the instrumental absorption.  The
detailed ISM edge structure closely follows the edge structure of
neutral oxygen as derived by theoretical R-matrix calculations.
However, the position of the theoretical edge requires a wavelength
shift. In addition the detailed instrumental RGS absorption edge
structure is presented. All results are verified by comparing to a
subset of Chandra LETG-HRC observations.


}

\maketitle

\section{Introduction}
In the past, detectors in the low energy (0.3 - 2.0 keV) X-ray band
offered spectra of modest spectral resolution. To properly model these
spectra, equivalent spectral resolution knowledge of the oxygen
K-edge absorption was required in order to correct the intrinsic source
spectra for the effects of interstellar extinction.  With the
appearance of high resolution X-ray spectrographs, notably the
XMM-Newton RGS (den Herder et al.  \cite{herder}) and Chandra LETG
(Brinkman et al. \cite{brinkman}), the ISM oxygen K-edge absorption can be
studied in much greater detail. These instruments, however, usually
suffer from absorption by oxygen located within the instruments. By
comparing spectra from sources with low and high extinction and
assuming that the interstellar cross-section per hydrogen atom is
constant over the sky, the instrumental and interstellar components can
be separated.

Models of extinction in X-rays by the
interstellar medium have been constructed, amongst others, by Morrison
and McCammon (\cite{morrison}). These models are based on assumed abundance's of
the ISM and cross sections by Henke et al. (\cite{henke}). However, the
spectral resolution in current spectrographs surpasses the resolution of
the Henke tables. 

\begin{figure}
  \resizebox{\hsize}{!}{\includegraphics[angle=90,clip]{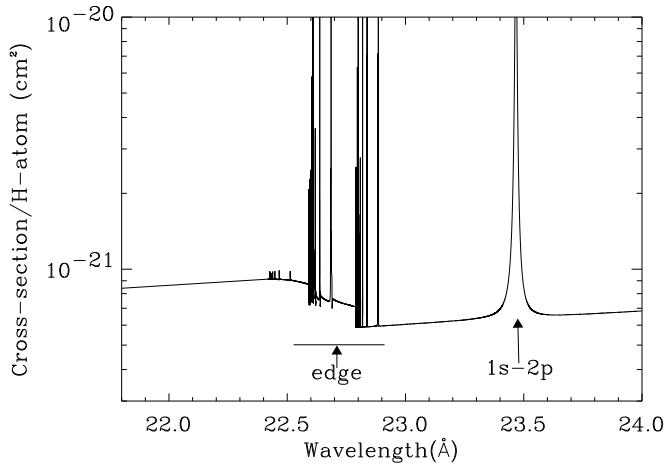}}
  \caption{Theoretical cross sections for the ISM around the oxygen 
	K edge based on the R-matrix (McLaughlin and Kirby, \cite{matrix}) 
	computations for neutral atomic oxygen combined with the 
         Morrison and McCammon (\cite{morrison}) profile for other elements.
	 The edge is not infinitely sharp, but consists of a series of
	 resonance features, indicated in the plot. The well known
	 1s-2p transition feature is also indicated.}
  \label{absmodel}
\end{figure}

For neutral atomic oxygen far more detailed absorption cross sections
have been calculated by the R-matrix approach (McLaughlin and Kirby,
\cite{matrix}). These cross sections include all the fine structure
resonance's along the K edge (around 22.6 - 22.9 \AA), including the
1s-2p transition at approximately 23.5 \AA.  Neutral oxygen is thought
to be the major constituent of the ISM. Therefore, a good model of the
detailed spectral structure around the oxygen K edge can be constructed
by combining the ``R-matrix" cross sections of neutral oxygen with the
Morrison and McCammon extinction cross sections for the remaining
elements. The ISM oxygen abundance as presented by Wilms et al.
 (\cite{wilms}) is used.  Fig.~\ref{absmodel} shows the combined
cross section.

In the matrix calculations uncertainties of up to 3.5 eV, or 0.15
\AA~are reported for certain energy levels which can lead to shifted
absolute positions of the resonance structures. The relative structure
however, which gives rise to the detailed shape of the observed edge,
can be expected to be known with much higher accuracy. The theoretical bound-bound
resonance's wavelengths (e.g. the 1s-2p line) were shifted to be in agreement with
laboratory data, such that for these transitions the error can be
expected to be less.

Of course interstellar oxygen can also be present in other forms then
neutral atoms. Ionised oxygen can be expected and oxygen can be
bound in molecules (e.g. OH, CO, $\rm{H}_2\rm{O}$), either as free
gases, or as a solid component of interstellar dust (e.g. ice,
silicates). Although the X-ray signature of ionised oxygen is well
known, the X-ray signature of bound oxygen is much harder to identify.
Depending on the detailed chemical and crystal structure and the
presence and structure of other compounds, the X-ray spectral
signatures may shift by up to 5 \% (chemical shift, see e.g. Sevier,
\cite{sevier}), and can be broadened. For this reason the exact X-ray
absorption spectrum of interstellar dust is not really known. Since
neutral atomic oxygen is believed to be the most significant constituent of
interstellar oxygen it is therefore justified to include its X-ray
absorption features only in our interstellar oxygen model.

In this paper we derive the shape of the interstellar oxygen absorption
edge by comparing the absorption features in sources with low and high
extinction.  We compare the neutral oxygen model cross sections with
the derived shape of the spectral profiles of interstellar extinction
observed by the RGS, corrected for the instrumental response.  Possible
deviations from the model will be discussed in terms of the presence or
upper limits of other oxygen components. In addition the detailed
features of the RGS instrumental absorption are resolved. A subset of
Chandra measurements is used to crosscheck the results obtained.

\section{Data}
To study spectral absorption features of the oxygen K edge, a
number of bright sources with simple intrinsic (power law) spectra
and no obvious narrow spectral features, were selected.
Table~\ref{sources} lists the sources. The corresponding
XMM-Newton orbits of the RGS data and Chandra observation ID's of the
LETG data are shown, as well as the $N_{\rm{H}}$ ISM column densities used. 

Since the main features in our spectra are caused by oxygen, we also show the
oxygen column density based on the ISM abundance by 
Wilms et al. (\cite{wilms}). For the extragalactic sources
(\object{Mrk~421}, \object{PKS~2155-304}, \object{3C~273}) we checked that we
got reasonable agreement with the oxygen spectral features, starting from the
$N_{\rm{H}}$ columns from an independent source (Kaastra, \cite{kaastra}),
while for the galactic sources (\object{Sco-X1} and \object{4U0614+091})
we actually fitted the oxygen column and convert to $N_{\rm{H}}$. It may be 
noted that for the derivation of the actual shape of the interstellar oxygen 
edge the precise oxygen column densities of the individual sources
are irrelevant.

Finally, for comparison only, we add the $N_{\rm{H}}$ column densities
based on \ion{H}{i} radio surveys. Since the line of sight towards the
galactic sources is likely to cross galactic molecular material, a
correction for the presence of $\rm{H}_2$ based on a ratio of
$\rm{H}_2$ to \ion{H}{i} of 0.3 (Dame et al. \cite{dame}) is applied.
For \object{Sco-X1} we assume all observed \ion{H}{i}
to be in front of the source, since its distance of 2.8 kpc (Bradshaw
et al. \cite{bradshaw}) and galactic latitude of
about 24 degrees places it well above the scale height of
$\approx$~150~pc of \ion{H}{i} in the galaxy.
For \object{4U~0614+091} a wide variety of distances is reported (see 
Brandt et al., \cite{brandt}) from 2 to 8~kpc although an
upper limit of 3~kpc seems likely. Combined 
with the low galactic latitude for this source of only 3 degrees, makes 
the estimated \ion{H}{i} column highly uncertain. 

Despite some uncertainties in the \ion{H}{i} column densities, 
table~\ref{sources} shows quite good agreement between our $N_{\rm{H}}$
columns derived from X-ray data and the \ion{H}{i} columns. This is a 
strong indication that the absorption is truly interstellar and not
originating at the source. This means the absorption characteristics 
can be assumed similar for all our sources.

On the RGS the oxygen edge is covered by CCD detector 4.
Unfortunately, early in the mission the CCD4 chain on RGS2 failed.  As
a consequence the oxygen data for on-axis sources originate from the
single CCD chip on RGS1 only. The bad spots (warm or hot pixels) on
this CCD detector 4 cause large systematic uncertainties in the spectra
so these regions have to be omitted from the analysis.  These bad spots
usually occur at the same wavelength for one source, but could be
slightly shifted for a different source, due to small differences in
source position with respect to the telescope axis.  By combining the
spectra of the low extinction sources \object{Mrk~421} and
\object{PKS~2155-304} the effect of these bad spots is reduced
significantly due to slight differences in the relative pointings of
these sources.

\begin{table*}
	\caption[]{Overview of sources and basic data}
	\label{sources}
	\begin{tabular}{lllllll}

	       & \multicolumn{2}{c} {Observation} & exposure$^\mathrm{i}$ & $N_{\rm{H}}$ column & 
	  	$N_{\rm{O}}$ column$^\mathrm{a}$ & $N_{\rm{H}}$ from \ion{H}{i} \\ 

	Source & XMM-Newton $^\mathrm{b}$ & Chandra $^\mathrm{c}$ &
		 (ksec) &
	       ($\times10^{20} \rm{cm}^{-2}$) &
	       ($\times10^{17} \rm{cm}^{-2}$) &
	       ($\times10^{20} \rm{cm}^{-2}$) \\

	\hline

	\object{Mrk~421}      &  84,171,259 &      & 172 & 
	                 $1.66 \pm 0.09 ^\mathrm{d}$ & $0.85 \pm 0.04$ & $1.61 \pm 0.10 ^\mathrm{e}$ \\

	\object{PKS~2155-304} & 87,174,362 & 3166 & 346 &
			 $1.25 \pm 0.08 ^\mathrm{d}$ & $0.65 \pm 0.04$ & $1.36 \pm 0.10 ^\mathrm{e}$ \\

	\object{3C~273}       & 277         &     & 90 &
			 $1.76 \pm 0.10 ^\mathrm{d}$ & $0.90 \pm 0.05$ & $1.68 \pm 0.10 ^\mathrm{e}$ \\

	\object{Sco-X1}       & 224,402 &         & 12 &
			 $20.0 \pm 1.5 ^\mathrm{f}$   & $10.2 \pm 0.8 $ & $19.0 \pm 3 ^\mathrm{g}$ \\

        \object{4U~0614+091}  & 231     & 100     & 23 &
			 $31. \pm 4. ^\mathrm{h}$    & $15.9 \pm 0.4 $ & $28 \pm 10 ^\mathrm{g}$ \\

	\end{tabular}
  \begin{list}{}{}
  \item[$^{\mathrm{a}}$] Derived from the $N_{\rm{H}}$ column depth based on
	                 the oxygen ISM abundance's from 
			 Wilms, Allen and McCray (\cite{wilms}) 
  \item[$^{\mathrm{b}}$] XMM-Newton orbit number.
  \item[$^{\mathrm{c}}$] Chandra Obs-ID.
  \item[$^{\mathrm{d}}$] Taken from independent Chandra data (Kaastra, \cite{kaastra})
  \item[$^{\mathrm{e}}$] Based on \ion{H}{i} observations by Lockman and Savage (\cite{lockman}).
  \item[$^{\mathrm{f}}$] Fitted on the RGS data (this paper). 
  \item[$^{\mathrm{g}}$] Based on the \ion{H}{i} survey by Hartmann and Burton (\cite{hartmann}) 
			 and a ratio of
			 $\rm{H}_2$ to \ion{H}{i} of 0.3 (Dame et al, \cite{dame}).
			 The errors are based on local gradients in the \ion{H}{i} map and
			 uncertainties in source distances.
  \item[$^{\mathrm{h}}$] Based on fits in both the Chandra and RGS data (this paper). 
  \item[$^{\mathrm{i}}$] Total RGS exposure time
\end{list}
\end{table*}

For the ``high extinction" source Sco-X1 a different approach was
adopted.  Sco-X1 is sufficiently bright to allow large offset pointings
(of about 27 arcmin.) which shifted the oxygen edge to CCD detector 3
(orbit 402).  Since both CCD3 detectors are active on both RGS's,
redundancy between the two CCD3 detectors solves the problem with
bad positions on the detector corresponding to a particular wavelength
range. Comparing the individual CCD3 spectra and an additional on-axis
pointing on Sco-X1 with the oxygen edge on CCD4 (orbit 224) it was
confirmed that the instrumental oxygen edge is identical for the
different detector chips, which implies that the same instrumental profile
applies irrespective of the location of the Oxygen edge on the instrument.
   
The data of \object{3C~273} were used as an independent check to verify that
the derived instrumental absorption profile for the RGS is correct. 

In all figures in this paper data are fluxed spectra, corrected
for exposure and the effective area excluding of course the effects of
instrumental oxygen absorption. Error bars represent the $1\sigma$
statistical error.

The lower effective area and limited exposure times of Chandra imply
higher statistical noise compared with XMM.  Including the Chandra
observations however, enables an independent check of the derived
properties of the interstellar oxygen absorption.  All Chandra data
shown are corrected for the overlapping $2^{\rm{nd}}$ and
$3^{\rm{rd}}$ orders.

\section{RGS analysis}

\begin{figure}
  \resizebox{\hsize}{!}{\includegraphics[clip]{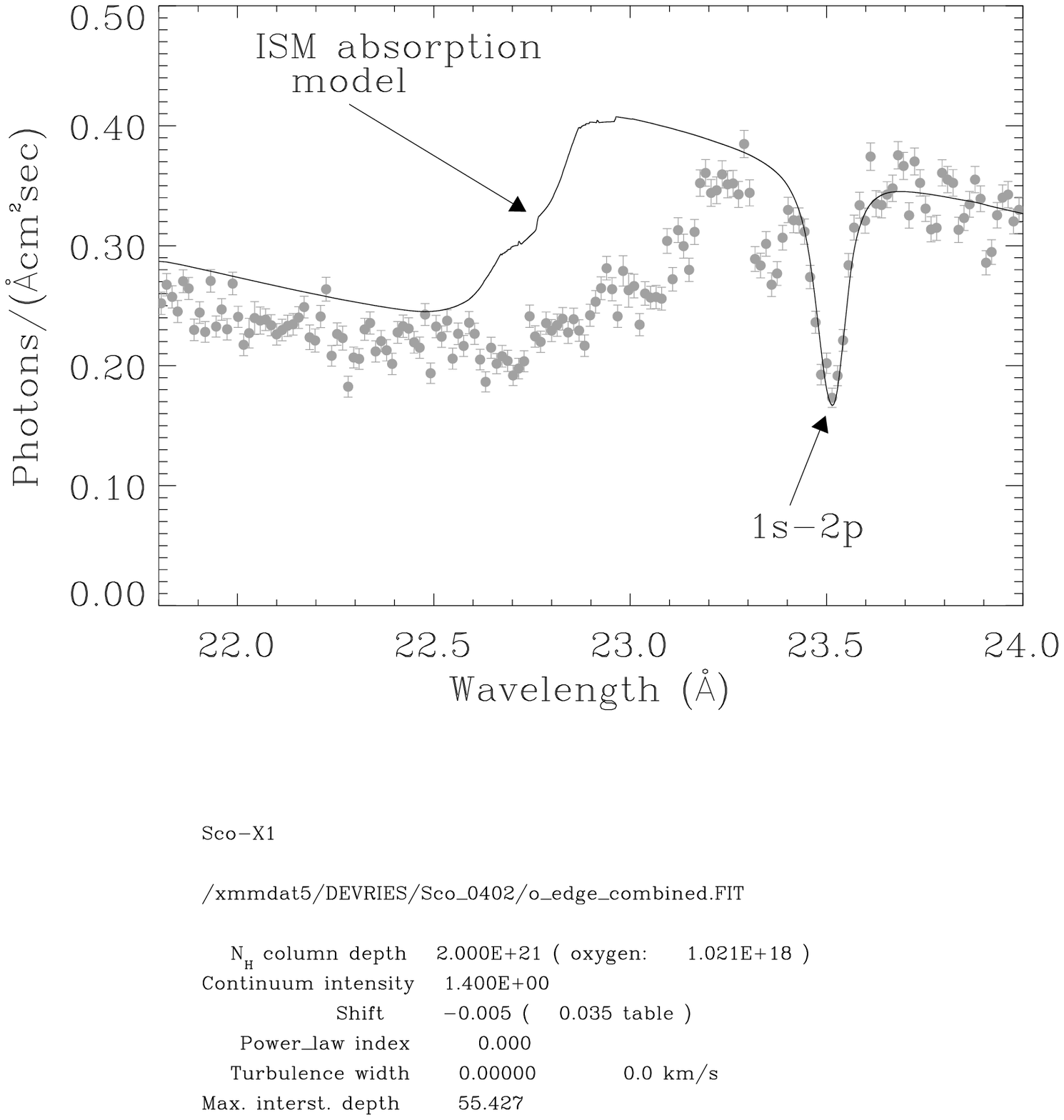}}
  \caption{RGS spectrum around the oxygen K edge of \object{Sco-X1} 
	   (grey data points). 
	   The solid black line represents the theoretical interstellar
           absorption (Fig.~\ref{absmodel}) convolved with the RGS instrumental
 	   resolution for an ISM column density of 
	   $N_{\rm{H}}=2.0\times10^{21} \rm{cm}^{-2}$. The ISM 1s-2p absorption
	   feature is also indicated.}
  \label{scox1}
\end{figure}

Fig.~\ref{scox1} shows the RGS spectrum around the oxygen K edge for
\object{Sco-X1}. Overplotted is the theoretical profile based on the
cross sections of  Fig.~\ref{absmodel}, convolved with the RGS
instrument response and using an extinction column depth of
$N_{\rm{H}}=2.0\times10^{21} ~\rm{cm}^{-2}$.  With this column depth the
1s-2p absorption line fits the data after applying a 30~m\AA~shift of
the theoretical position to longer wavelengths.  This is somewhat
larger than expected based on the known accuracy of the RGS wavelength
scale of $1\sigma = 8$~m\AA~(den Herder et al., \cite{herder}). 

\begin{figure}
  \resizebox{\hsize}{!}{\includegraphics[angle=-90,clip]{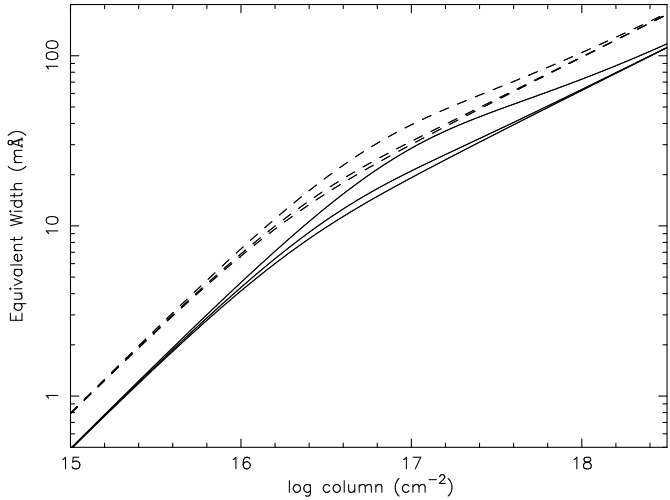}}
  \caption{Relation between equivalent line width of the 1s-2p absorption and 
	   $N_{\rm{O}}$ column density. The solid lines show the curves for neutral oxygen
	   for different velocity dispersions of 10, 30 and 100 km/s from the lowest
	   to the highest solid line respectively. 
	   Curves for velocities $\leq$ 10 km/s overlap with the lowest curve (10 km/s). 
	   The broken lines show the same for singly ionised oxygen (\ion{O}{ii}). 
	  }
  \label{kol}
\end{figure}
 
Due to the very high optical depth of the narrow 1s-2p line core, saturation
effects can be expected. Given the cross sections of the 1s-2p line as shown
in Fig.~\ref{absmodel}, the expected relation between equivalent width of the line
and the column density was computed for a variety of velocity dispersions.
This is shown in Fig.~\ref{kol}. The bend in the curves for high column
density does indicate saturation effects, but due to the broad wings in
the line (contrary to e.g. lines of \ion{O}{i} in the UV), the equivalent 
width of the line keeps increasing steadily even for high optical depths. 
This means that even for high extinction the column density can be reasonably
well determined from the equivalent width.
However, the actual equivalent width and  
 depth of the observed line profile do depend on the velocity
dispersion along the line of sight. Entering realistic velocity
dispersions ($<$ 200 km/s) did not improve the fit sufficiently to
allow for tight constraints on the velocity dispersion and column
depth. Therefore a fit of the 1s-2p line only, leaves a rather large
uncertainty on the actual column density.
Fixing the column depth to $N_{\rm{H}}=2.0\times10^{21} ~\rm{cm}^{-2}$
and performing more detailed modeling of the 1s-2p line depth using the
SPEX package (Kaastra, \cite{kaastra3}) points to a velocity dispersion
in the range of 30 - 50 km/s.
However, this is highly dependent on the
real distribution of the velocity components. In addition, given the errors on the
equivalent widths (Table~\ref{features}) and too poor resolution of the
RGS for this purpose, the fitted velocity dispersion can only be
seen as an indication only.

Since the depth of the absorption below the edge (below 22.5~\AA) is a
measure for the total of ISM and instrumental extinction, it was
checked whether the difference between ISM absorption model and data below
the edge can be attributed to the instrument. The depth of the
instrumental oxygen absorption determined later in this paper, does
indeed match this offset. Since the depth of the absorption below the
edge extends over a large wavelength range and is therefore not
influenced by ISM velocity dispersions, it is believed that
the current fit does give a
good measure of the true oxygen column depth for this source. The
derived column density is also in good agreement with the numbers
retrieved from \ion{H}{i} observations (see Table \ref{sources}).

\begin{figure}
  \resizebox{\hsize}{!}{\includegraphics[clip]{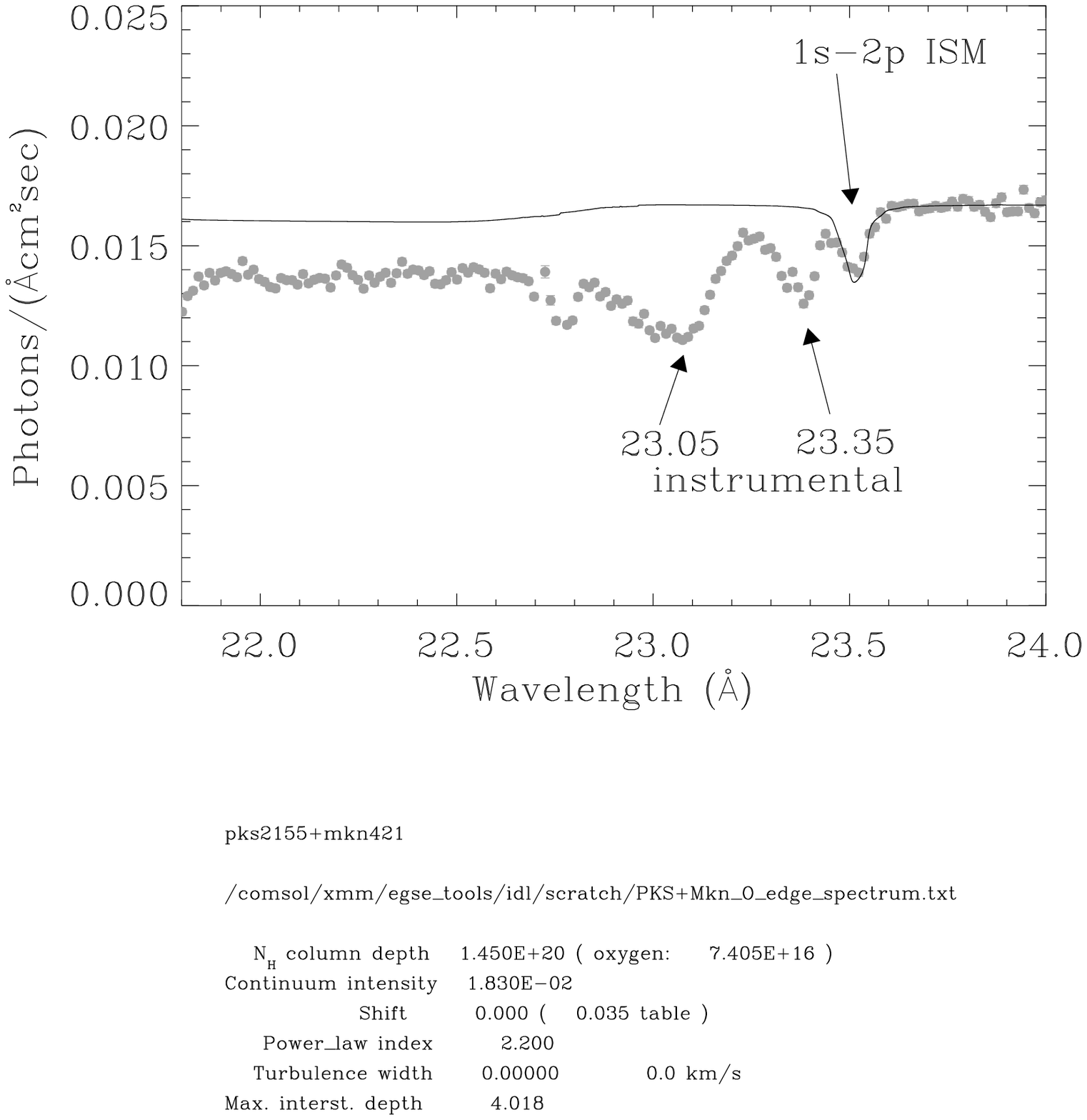}}
  \caption{RGS combined spectrum around the oxygen K edge for the low extinction
	  sources (\object{Mrk~421}, \object{PKS~2155-304}). A fit of the theoretical ISM
	  absorption model to the 1s-2p line at 23.5~\AA~is also shown
	  (black line). Main difference with the data points is due to the 
	  additional instrumental absorption. Two key instrumental features at
	  23.05 and 23.35~\AA~are indicated. 
 }
  \label{mkn}
\end{figure}

Fig.~\ref{mkn} shows the RGS spectrum for the sum of \object{Mrk~421}
and \object{PKS~2155-304} data.  This is the ``low extinction" data.  Of
course this summed spectrum yields a kind of weighted average of the
spectral characteristics of the two individual objects \object{Mrk~421}
and \object{PKS~2155-304}. Since we are not interested in the intrinsic
source spectra the large scale shape of the combined spectrum is not
relevant. The $N_{\rm{H}}$ column depth for both sources is low and
does not differ too much between the sources. It is therefore estimated
that the distortion of the ISM absorption profile due to adding the two
sources is well within the statistical uncertainty of the individual 
data points.

For these ``low extinction" data, the model wavelengths are shifted by
35~m\AA~to longer wavelengths to fit the 1s-2p feature. This shift is
consistent with the 30~m\AA~shift in Fig.~\ref{scox1} given the 8~m\AA~
error on the RGS wavelength scale. Main difference between the
data points of these low extinction sources and the overplotted ISM
absorption is due to the additional instrumental absorption. Two key
instrumental features can be clearly recognised at around 23.05 and
23.35~\AA.  This is the fingerprint of the actual compounds in the
instrument which cause the absorption.  These dips can also be noticed
in Fig.~\ref{scox1}, in the \object{Sco-X1} data.  The fact that the
equivalent width of the 23.35 feature (see Table~\ref{features}) is the
same for both Sco-X1 and the low extinction sources, despite the fact
that the $N_{\rm{H}}$ column densities differ more than a factor of 10,
does confirm the instrumental nature of this feature.

\begin{figure}
  \resizebox{\hsize}{!}{\includegraphics[angle=90,clip]{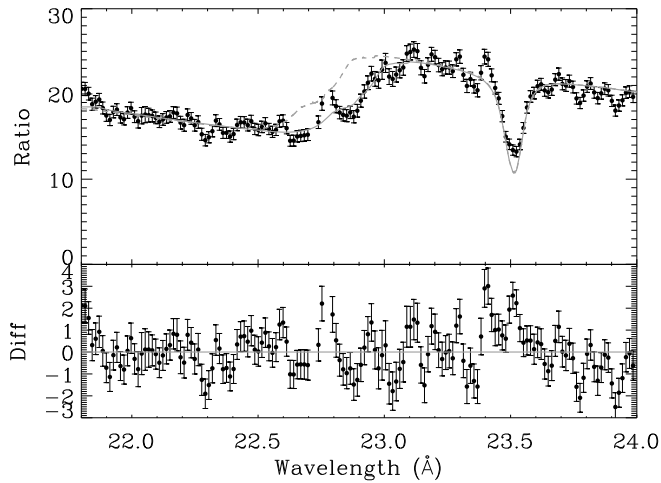}}
  \caption{ Ratio of the spectra of Figs.~\ref{scox1} and~\ref{mkn} (black
	    data points). 
	    This represents the variable interstellar extinction between the
	    sources. The broken grey line shows the fitted theoretical 
	    extinction. The solid grey line shows this theoretical extinction
	    with the edge shifted by 0.13 \AA. Error bars shown represent
	    the $1\sigma$ statistical error.}
  \label{ratio}
\end{figure}

Taking the ratio between the ``low" and ``high" extinction data will
remove all the features which are constant between the data, in
particular the instrumental features. What is left must represent the
true interstellar extinction. This ratio is shown in Fig.~\ref{ratio}
(black data points).

The theoretical ISM extinction curve (broken grey line) is fitted to
the depth of the edge below 22.5~\AA. By comparing the theoretical
oxygen edge with the extinction data, it appears that the shape of the
edge matches the data, but that the position of the edge is shifted.
The solid grey line shows this effect. It is identical to the broken
line but with the edge (all the resonance structures identified with
the indicated edge in Fig.~\ref{absmodel}) shifted by 0.13~\AA, to
larger wavelengths.  The error uncertainty of this shift is estimated
at 0.02~\AA.  Hereafter, we use these absorption cross sections with shifted
oxygen edge as our model of the interstellar
oxygen K edge.

It shows clearly in Fig.~\ref{ratio} that the absorption dips around
23.05 and 23.35~\AA, which are present in both the low and high
extinction data (Figs.~\ref{scox1} and~\ref{mkn}) have largely disappeared in
the ratio plot, which indicates that both features are generated by
the RGS instrument itself. Possible residues deviate less than $3\sigma$ from
the model.

Any deviations in Fig.~\ref{ratio} from the model can point to
absorption features of other components, in particular other forms of
oxygen. Any amount of oxygen, either atomic, in molecules or solid
compounds will be reflected by the depth of the oxygen edge. Therefore
the depth of the edge by itself cannot separate different components.
Oxygen compounds other then the neutral atomic form can only be seen if
they show distinct, sufficiently narrow, absorption features. In order to judge
possible strengths of other absorption features, the level of the
fitted continuum is important. In Fig.~\ref{ratio} it shows that the
continuum is very well constrained by the data below 22.5~\AA~and above
23.7~\AA. Although fit residues between 22.5 and
23.5~\AA~seem to suggest additional structure, the tight constraints
on the continuum level do not allow much freedom to move the continuum
up to a higher level such that there is room for clear absorptions.
Taking into account the error on the fitted continuum, the fit residues
can be analysed in terms of upper limits on possible other oxygen
compounds.

\begin{figure}
  \resizebox{\hsize}{!}{\includegraphics[clip]{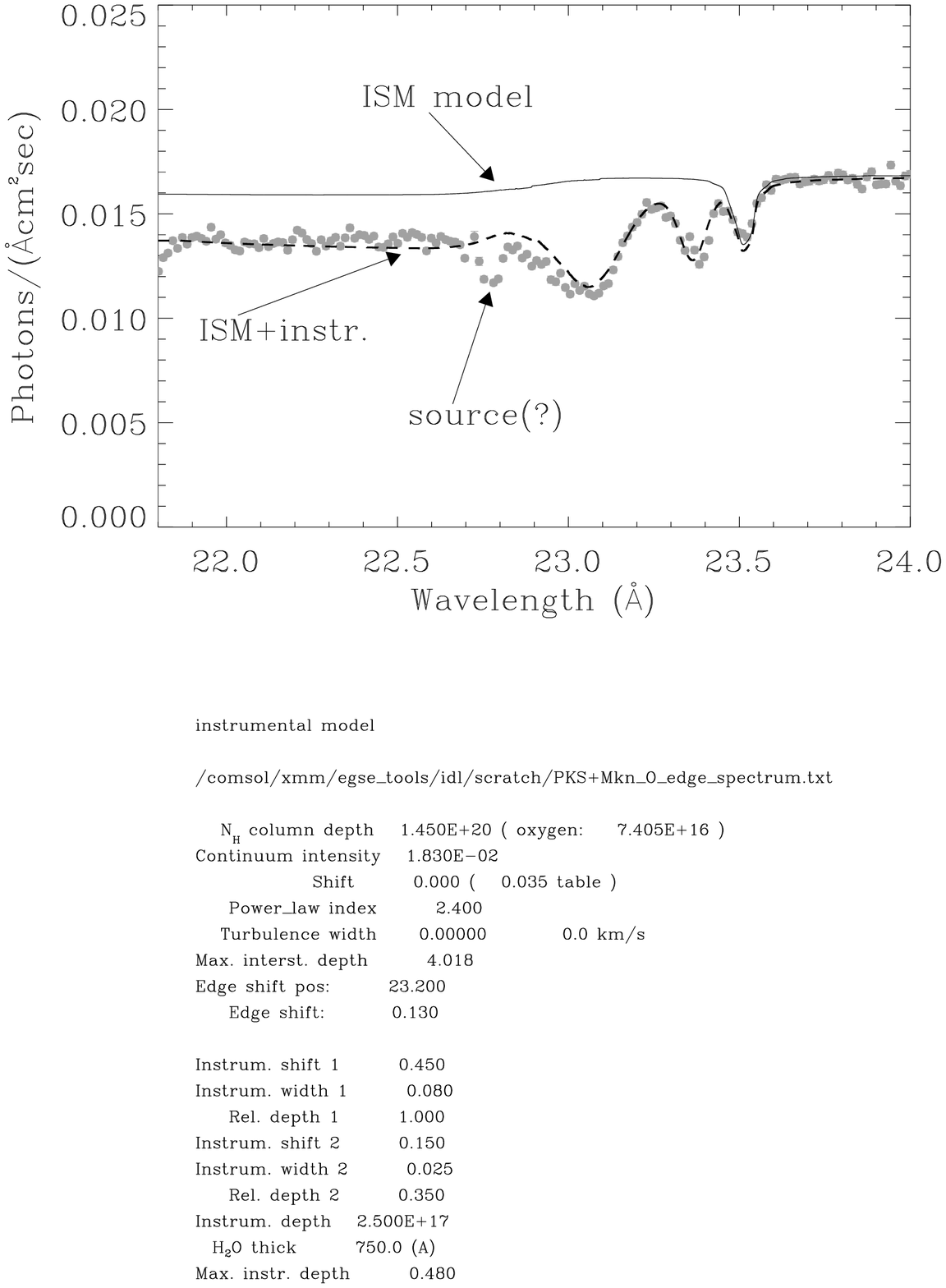}}
  \caption{ Fit of modified 1s-2p lines to the instrumental
  	    features for the low extinction data (Fig.~\ref{mkn}).
	    The broken line shows the combination of our interstellar 
	    extinction model (black line) and the fit of a broadened
	    shifted atomic 1s-2p line model to the 23.05 and
	    23.35~\AA~instrumental 
	    features. }
  \label{ffit}
\end{figure}

A striking feature in the fit residue is the feature around 22.77~\AA.
At this wavelength some excess flux appears in the ratio plot
(Fig.~\ref{ratio}), which means that this feature is not present in the
\object{Sco-X1} data (Fig.~\ref{scox1}). Inspection of the individual
XMM data show that this feature is present in both \object{Mrk~421} and
\object{PKS~2155-304} data but not in the 4U~0614+91 and
\object{Sco-X1} spectra. The spectrum of the combined low extinction
sources, Fig.~\ref{ffit} shows this feature as a clear distinct
absorption feature. It is unlikely that the apparent excess flux in the
ratio plot is caused by two absorption dips on both sides of the
22.77~\AA~feature, e.g. due to solid ISM oxygen compounds. In that case
the \object{Sco-X1} should have shown the 22.77~\AA~feature in
``emission" as well. In the subsequent section we will show that the
Chandra data of \object{PKS~2155-304} and 4U~0614+91 show the same
behaviour as the RGS data. For this reason this feature is attributed to
the \object{PKS~2155-304} and \object{Mrk~421} source spectra. 

If we tentatively identify this feature (Fig.~\ref{ffit}) in the 
\object{PKS~2155-304} and \object{Mrk~421} spectra with an \ion{O}{iv} 
absorption blend at 22.75~\AA, given the
equivalent width of $\approx 10$m\AA, we derive a column density of 
$N_{\ion{O}{iv}}$ of $6 \times 10^{15} \rm{cm}^{-2}$. The most likely
candidate for this absorption is the local filament as observed e.g. in 
\ion{O}{vii} by Nicastro et. al. (\cite{nicastro}), with a column density
of $4 \times 10^{15} \rm{cm}^{-2}$. However, since Nicastro et. al. derive
an \ion{O}{vi} column of only $0.1 \times 10^{15} \rm{cm}^{-2}$, the 
\ion{O}{iv} is most likely due to a less ionised component. If the gas is
photo ionised, we expect \ion{O}{iv} to peak at $\rm{log}\xi \simeq -0.8$.
For this ionization parameter the column density of the neighbouring
\ion{O}{iii} and \ion{O}{v} ions is a factor of 3 smaller than the 
\ion{O}{iv} column and we estimate the strongest spectral features 
of those ions to be at 22.33~\AA (\ion{O}{v}, eq. width = 4 m\AA)
and 23.08~\AA (\ion{O}{iii}, eq. width = 2 m\AA). Equivalent widths
of this order of magnitude are not completely excluded by the data, taking
into account that these lines positions are uncertain by about
$\approx 0.03$~\AA. Unfortunately, there is no clear positive indication
of these lines.
   
Looking in detail to other features in the fit residues of the ratio
spectrum of Fig.~\ref{ratio}, features can be recognized at 23.05 and
23.35~\AA. The 23.05~\AA~feature seems an
incomplete cancellation of the 23.05~\AA~instrumental feature and has an
equivalent width of $3\pm3$~m\AA. Finally there seems some structure around
23.35~\AA~which has an equivalent width of $4\pm3$~m\AA. 
A tentative explanation could be traces of \ion{O}{iii} (at 23.05~\AA) and 
\ion{O}{ii} (at 23.35~\AA). This implies column densities of 
$\approx 5 \times 10^{15} \rm{cm}^{-2}$ (see Fig.~\ref{kol}). All these column
densities are sufficiently low, such that their combined K-edge has an optical
depth less than 0.01.
This puts constraints on the average ionization stage of oxygen
in the interstellar medium. Higher ionization stages of Oxygen may dominate
the hotter components of the interstellar medium.         

Apart from possible lines of ionised oxygen, oxygen locked in solids
may cause features in the fit residues of Fig.~\ref{ratio}.
It is well known that in solids the spectral features of X-ray
absorption lines can be shifted (see e.g. Sevier, \cite{sevier}) and
broadened with respect to the corresponding lines in free atoms.
For broader lines saturation effects are less critical.
If the combined features identified above would be attributed to 
shifted and broadened 1s-2p lines of neutral oxygen
in solid compounds, we estimate the amount of oxygen in these features at 
less than 10~\% of the neutral oxygen.
In case of possible saturation effects the estimated upper limits will become
smaller. If however the combined chemical shifts of lines of oxygen locked in solids
smear out its signature over a broad wavelength band, its signature will be completely
lost and large amounts of oxygen may be hidden in solids which are simply not
detected by X-ray absorption features.     

The fact that the depth of the 1s-2p line does not quite match the data
is not really relevant.  As explained before, the relative depths of
the edge and the line of the high extinction component in the ratio
data is rather arbitrary and depends on other factors (ISM velocity
field) then just the column depth of the absorbing material.

If the
instrumental features at 23.05 and 23.35~\AA~are interpreted as shifted
1s-2p lines of bound oxygen in some solid compounds located on the
detectors or optics, the column density for these features can be
fitted.  First our interstellar extinction model is fitted to the
interstellar 1s-2p line of the low extinction data at 23.5 \AA
(Fig.~\ref{ffit}). 
Fig.~\ref{kol} shows that for the fitted ISM column density of
$N_{\rm{H}} = 1.5~\times~10^{20}~\rm{cm^{-2}}$, which corresponds to
$N_{\rm{O}} = 0.76~\times~10^{17}~\rm{cm^{-2}}$, line saturation effects
are minor and only uncertainty in velocity dispersion will influence the
determined column density. We assumed velocity dispersions $\leq 10$ km/s
for these high galactic latitude sources.  

In addition to this ISM absorption the
instrumental 23.05 and 23.35~\AA~features are fitted by shifting and
broadening the 1s-2p line.  It now appears that with this instrument
model the depth of the edge below 22.5~\AA~is also well fitted.
This is interpreted as a confirmation that the 23.05
and 23.35~\AA~features are indeed shifted and broadened 1s-2p lines.
The oxygen column density in both instrumental features amounts to
$N_{\rm{O}}=2.5\pm0.2~\times10^{17}~\rm{cm}^{-2}$.  If all this
instrumental oxygen would be in the form of water(ice) only, this
column density amounts to a layer of about $750\pm50$~\AA~thickness. However,
the fact that we see two features might indicate that the oxygen is
present in two different compounds.

\begin{figure}
  \resizebox{\hsize}{!}{\includegraphics[angle=90,clip]{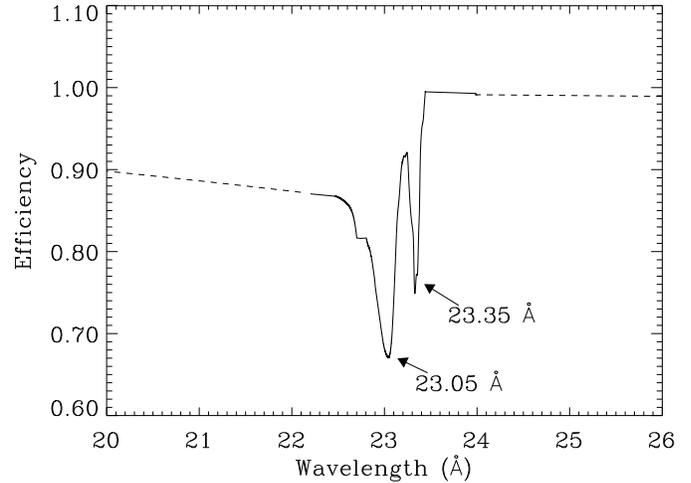}}
  \caption{ The derived instrumental efficiency due to the instrumental oxygen 
	    absorption. The extrapolations beyond the wavelength range
	    of 21.8 - 24.0~\AA~have been derived from the 
	    Henke~(\cite{henke}) tables (broken lines). }
  \label{eff}
\end{figure}

The actual difference in Fig.~\ref{ffit} between the fitted theoretical
interstellar extinction curve and the data (ignoring the 22.77~\AA~
feature) can be interpreted as the instrumental oxygen absorption. For
these low extinction sources, any uncertainty in the exact shape of the
interstellar oxygen edge, whether or not features of other forms of
oxygen then just neutral oxygen are present, hardly plays a role. The
derived corresponding efficiency of the RGS instrument is plotted in
Fig.~\ref{eff}. This efficiency is implemented in the current
calibration files (CCF) for the RGS response, used by the XMM SAS data
reduction software package (Jansen et al. \cite{jansen}). Note that
some of the sharp ``edges" in the graph result from the fitting process
used to derive the instrumental curves.  They have no physical
meaning.  These sharp features disappear when this correction is applied to
the instrument response.

\begin{figure}
  \resizebox{\hsize}{!}{\includegraphics[clip]{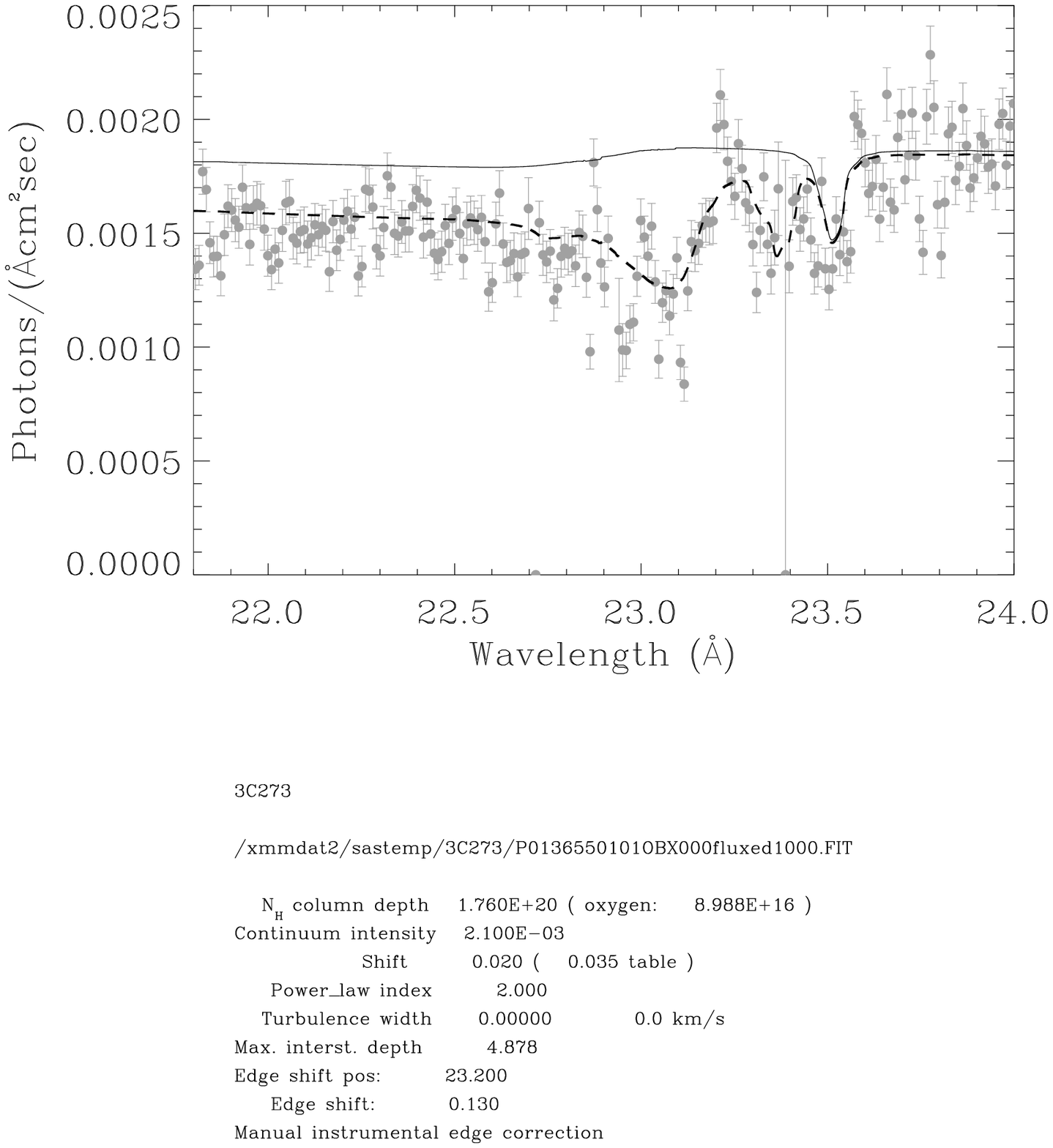}}
  \caption{Fit of our interstellar model with a column depth of
	   $N_{\rm{H}}=1.76\times10^{20} \rm{cm}^{-2}$ and our instrumental
	   efficiency model to the RGS spectrum of \object{3C~273} in the 
	   wavelength range around the oxygen K edge }
  \label{3c273}
\end{figure}

Our interstellar and instrumental absorption models around the oxygen K
edge were verified by applying them to
\object{3C~273}.  The statistical and systematic (due to warm/hot
columns in the RGS CCDs) noise is higher for this data than the
datasets used above.  Using
a column depth of $N_{\rm{H}}=1.76\times10^{20}~\rm{cm}^{-2}$, derived
from independent Chandra data, yields
an acceptable fit to these data (Fig.~\ref{3c273}).

\section{Cross checking on Chandra data}

The method to compare data from low and high extinction sources to
derive absorption characteristics of the ISM can also be applied to
Chandra data.  We looked at a subset of Chandra LETG HRC data, since
this Chandra grating/detector combination has the highest effective area
at the oxygen K-edge. The observations are listed in Table~\ref{sources}.

\begin{figure}
  \resizebox{\hsize}{!}{\includegraphics[clip]{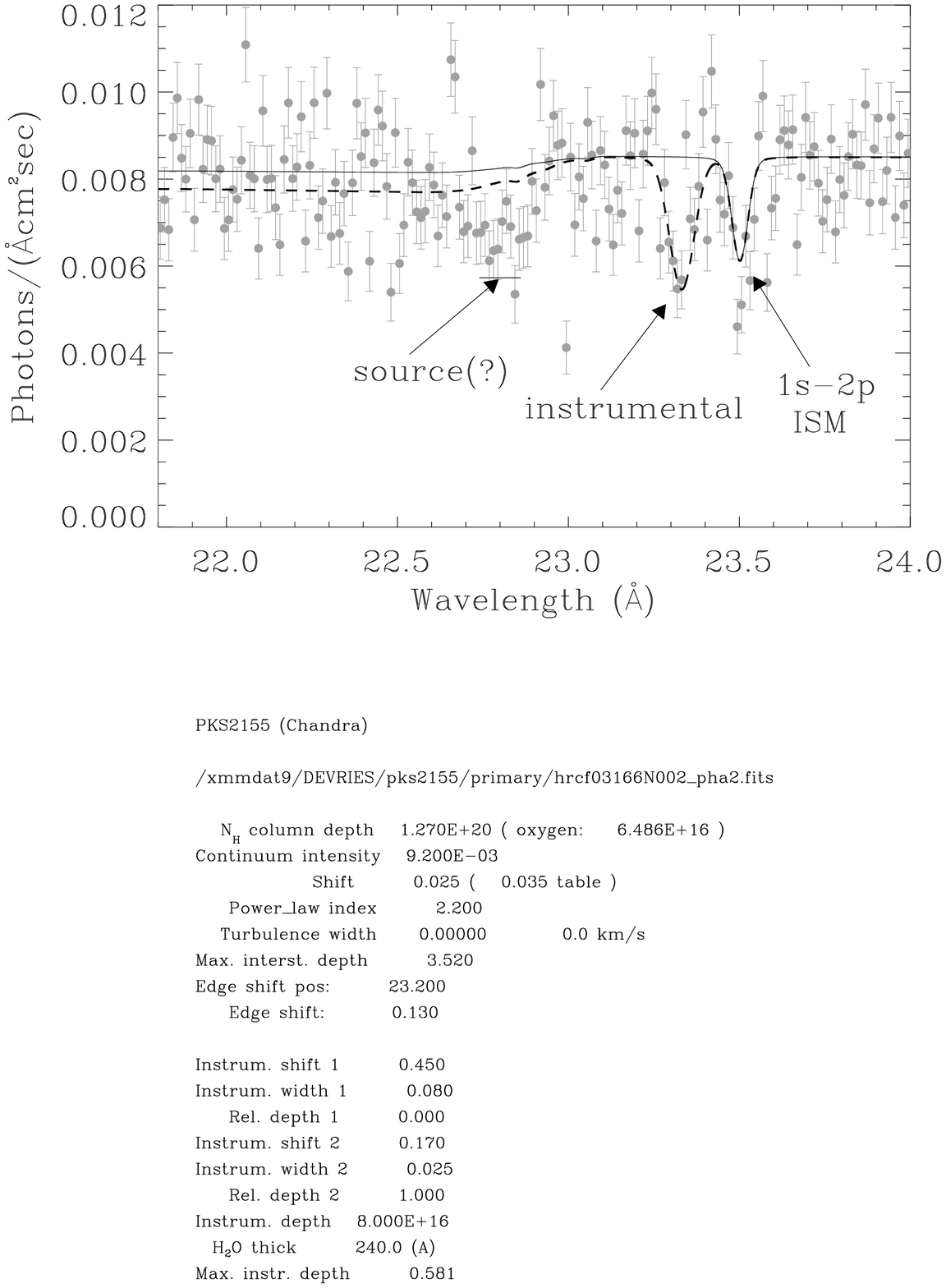}}
  \caption{Chandra LETG HRC observation of \object{PKS~2155-304}. The solid black line
  	   represents the ISM absorption using our 
	   $N_{\rm{H}}$ of $1.25\times10^{20} \rm{cm}^{-2}$. The broken line shows 
	   a fit to the instrumental line at 23.35~\AA~which can also be
	   recognised in the XMM-RGS data. Here, this fit corresponds to an
	   oxygen column depth of $8\times10^{16} \rm{cm}^{-2}$. 
	   The feature at 22.77~\AA, which is also seen in the XMM-RGS
           is attributed to the source itself.
 }
   \label{pks2155_chandra}
\end{figure}

Fig.~\ref{pks2155_chandra} shows the Chandra observation of the low
extinction source \object{PKS~2155-304}. Comparing with the XMM-RGS observation
of the low extinction sources in Fig.~\ref{mkn} it is immediately
apparent that the deep instrumental absorption feature seen at 23.05~\AA~in
the XMM data is largely absent in the Chandra data. The
23.35~\AA~instrumental absorption feature however can be recognised in
the Chandra data. This feature is treated similarly as for the XMM-RGS,
by assuming it to be a shifted/broadened instrumental 1s-2p line. This
fit is shown with the plotted broken line. Also the suspected
absorption due to the source at 22.77~\AA~can be recognised in the Chandra
data.

From these observations it follows that the general instrumental absorption
characteristics between Chandra-LETG-HRC and XMM-RGS differ, but
the 23.35~\AA~component is common in both instruments. 

\begin{figure}
  \vbox{
  \resizebox{\hsize}{!}{
     		\includegraphics[clip]{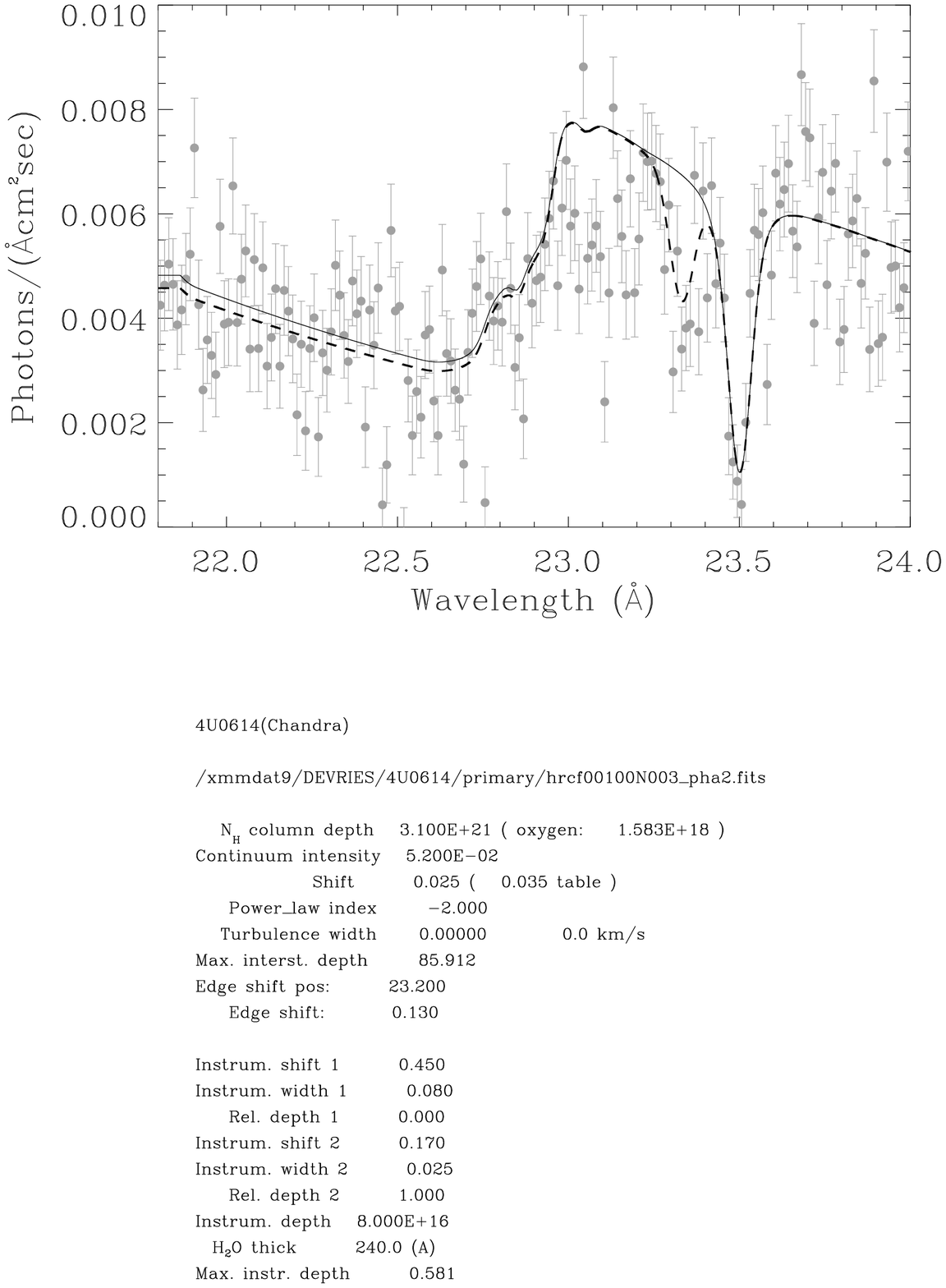}}
  	     \caption{Chandra LETG HRC observation of \object{4U~0614+091}.
		      The solid 
		      line shows the modeled interstellar absorption using
		      the Chandra wavelength resolution, while the
		      broken line adds the instrumental absorption as
		      determined in the \object{PKS~2155-304} observation 
			(Fig.~\ref{pks2155_chandra}). }
  	     \label{4u0614_chandra}
  \vspace{0.1in}
  \resizebox{\hsize}{!}{
     		\includegraphics[clip]{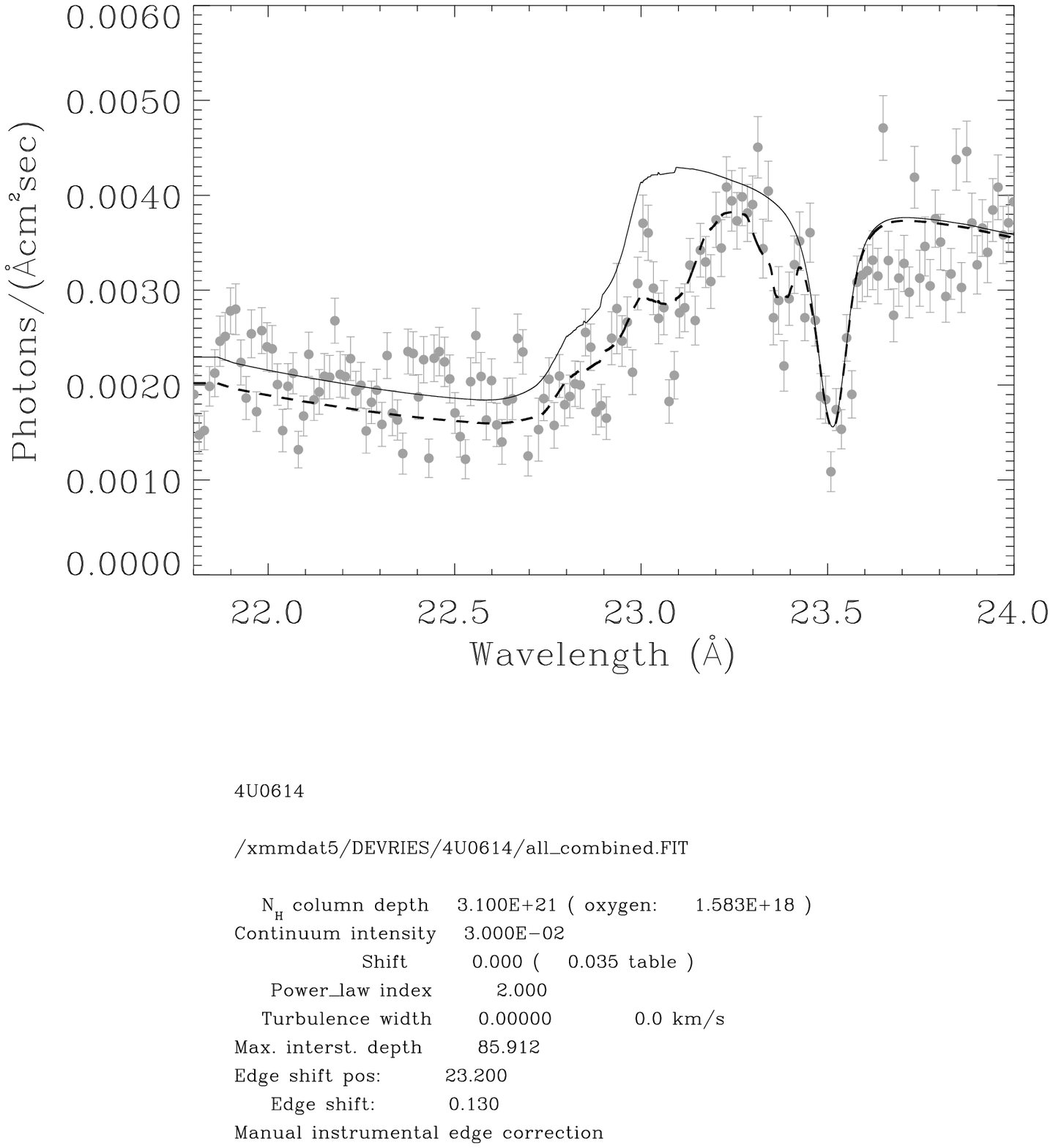}}
             \caption{XMM-RGS observation of 4U~0614+91. These data can be 
	        directly compared with Fig.~\ref{4u0614_chandra}. The
		broken line adds the RGS instrumental absorption. It can be seen
		that the 1s-2p line in the Chandra data shows deeper absorption
		due to the higher spectral resolution of Chandra.}
             \label{4u0614_xmm}
  }
\end{figure}

A high extinction source observed by Chandra is \object{4U~0614+091}
(Paerels et al., ~\cite{paerels}). The Chandra spectrum of this source
is shown in Fig.~\ref{4u0614_chandra}. The corresponding XMM-RGS
spectrum is shown in Fig.~\ref{4u0614_xmm}.  It can be noted that the
Chandra 1s-2p line shows a deeper absorption than the XMM-RGS
spectrum.  This is due to the higher spectral resolution of Chandra of
$\Delta\lambda=0.045$ FWHM (Chandra Proposers Observatory Guide)
compared to $\Delta\lambda=0.065$ FWHM (den Herder et al.
\cite{herder}) for the RGS and means that this line is not resolved by
the XMM-RGS. 

Both \object{PKS~2155-304} and \object{4U~0614+091} Chandra data
show an apparent wavelength inconsistency of 50~m\AA~to larger
wavelengths compared to the
theoretical position of the 1s-2p line.  This can be compared with the
30 to 35~m\AA~shift in the same direction for the XMM-RGS data noted
before. The systematic wavelength error on the Chandra data is
estimated at 10~m\AA~in our wavelength region (Kaastra et al.,
\cite{kaastra2}), while we mentioned a wavelength inaccuracy of
8~m\AA~for the RGS before.  This means that, within the errors
reported, the Chandra and RGS wavelength scales match, but that likely
the R-matrix theoretical 1s-2p line position requires a slight
adaptation of $35\pm10$~m\AA~to longer wavelengths.  This can be
compared with the noted discrepancies between the theoretical R-matrix
computations and laboratory measurements of $\approx$0.25 \% or
60~m\AA. Laboratory measurements report wavelengths between 23.49~\AA~
(Menzel et al., \cite{menzel}) and 23.53~\AA~ (Stolte et al.,
\cite{stolte}).

In fig.~\ref{4u0614_chandra} the broken black line shows the fitted
model spectrum corrected for the chandra instrumental absorption as
derived from fig.~ \ref{pks2155_chandra}. There is a hint for some
additional (instrumental) absorption between 23.0 - 23.2 \AA,
although the \object{PKS~2155-304} data do not show this absorption
that clearly. However, the Chandra data are quite noisy in this respect.

\begin{figure}
  \resizebox{\hsize}{!}{\includegraphics[angle=90,clip]
			   {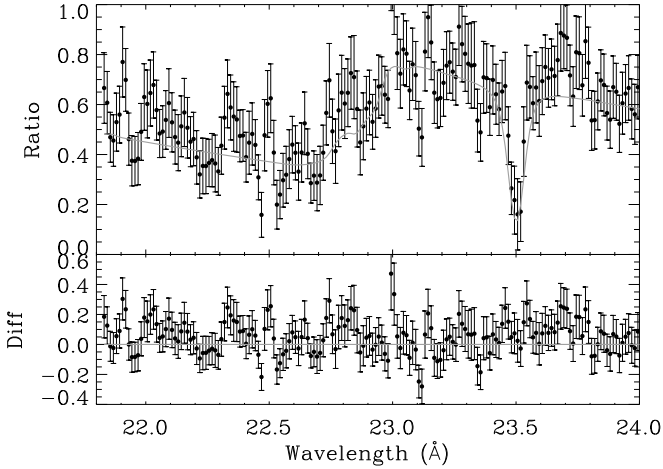}}
  \caption{ Ratio of the spectra of Figs.~\ref{4u0614_chandra} 
	    and~\ref{pks2155_chandra} (black data points). 
	    This shows the interstellar extinction according to Chandra.
	    Overplotted (grey line) is our interstellar extinction model
	    which includes the 0.13~\AA~shift of the oxygen edge.
	    }
  \label{interstellar_chandra}
\end{figure}
   
Combining the Chandra spectra of Figs.~\ref{4u0614_chandra}
and \ref{pks2155_chandra} yields the interstellar extinction curve
according to Chandra, shown in Fig.~\ref{interstellar_chandra}. Since
all instrumental features are removed from this plot, it can be
directly compared with the corresponding XMM-RGS plot of
Fig.~\ref{ratio}.  The solid grey line in both plots shows our ISM
absorption model which is identical in both Fig.~\ref{ratio} and
Fig.~\ref{interstellar_chandra}.  Although the Chandra profile is much
noisier, the profiles are remarkably
identical. Even the positive excursions around 22.77~\AA~which hint at
source absorption features, are repeated in the Chandra data although
with far less significance. This similarity between the XMM-RGS and
Chandra derived interstellar oxygen absorption features confirms the
validity of the data analysis.

\begin{table*}
  \caption[]{Summary of spectral features around the neutral oxygen-K edge}
  \label{features}
  \begin{tabular}{cllllll}
  wavelength &      & \multicolumn{5}{c} {Observed equivalent width$^{\mathrm{a}}$ (m\AA) in source spectrum} \\
    (\AA)    & identification & Mrk421 & PKS 2155 & 3C273 & Sco-X1 & 4U 0614 \\
  \hline

  $23.50 \pm 0.01 ^\mathrm{b}$ & \parbox[t]{1.5in}{neutral oxygen 1s-2p interstellar absorption line} &
	 $16 \pm 3$ & \parbox[t]{0.5in}{ $15 \pm 3$ \\ $20 \pm 5^{\mathrm{d}}$ }  & $20 \pm 8$ & $60 \pm 5$ & 
		      \parbox[t]{0.5in}{ $70 \pm 8$ \\ $80 \pm 5^{\mathrm{d}}$ }
   \vspace{6pt} \\

  $22.6 - 23.0^{\mathrm{e}}$ & \parbox[t]{1.5in}{interstellar oxygen-K edge absorption structure} &
            $)^{\rm{f}}$ &  $)^{\rm{f}}$ & $)^{\rm{f}}$ & $)^{\rm{f}}$ &$)^{\rm{f}}$            
  \vspace{6pt} \\

  $23.05 \pm 0.08^\mathrm{c}$ & \parbox[t]{1.5in}{instrumental absorption} &
            $)^{\rm{g}}$ &  $)^{\rm{g}}$ & $)^{\rm{g}}$ & $)^{\rm{g}}$ &$)^{\rm{g}}$            
  \vspace{6pt} \\

  $23.35 \pm 0.02$ & \parbox[t]{1.5in}{instrumental absorption} & 
	  $18 \pm 4$ & \parbox[t]{0.5in}{ $15 \pm 4$ \\ $25 \pm 5^{\mathrm{d}}$ }  & $18 \pm 6$ & $19 \pm 4$ & 
		       \parbox[t]{0.5in}{ $19 \pm 3$ \\ $30 \pm 7^{\mathrm{d}}$ }
  \vspace{6pt} \\

  $22.77 \pm 0.03$ & \parbox[t]{1.5in}{possible local intergalactic absorption} &
          $ 9 \pm 2$ &                    $11 \pm 6$                               & \textless 5  & \textless 3 &
		       \textless 5 
  \vspace{6pt} \\

  \end{tabular}
\begin{list}{}{}
\item[$^{\mathrm{a}}$] Equivalent width of feature in the RGS spectrum, unless noted otherwise.
\item[$^{\mathrm{b}}$] Shifted by $35 \pm 10$ ~m\AA~to longer wavelengths with respect to the theoretical position
		       of McLaughin and Kirby, (\cite{matrix}).
\item[$^{\mathrm{c}}$] Broad structure of Oxygen bound in a solid compound. The error also reflects
		       the broad and asymmetric nature of the feature.
\item[$^{\mathrm{d}}$] Equivalent width of feature in the Chandra spectrum.
\item[$^{\mathrm{e}}$] The K edge is not sharp, but extends over the indicated range. The edge is shifted by $0.13\pm0.02$~\AA with respect to the theoretical
R-matrix calculations.
\item[$^{\mathrm{f}}$] An equivalent width has no meaning for an edge
		       structure.
\item[$^{\mathrm{g}}$] Since this instrumental feature is very broad the
 continuum level and structure is very uncertain. For this reason no
 equivalent width was determined.
\end{list}
\end{table*}
 
\section{Discussion and conclusions}
Comparing high resolution X-ray spectra of sources with low
and high interstellar extinction, absorption features around the oxygen K edge
can be separated into instrumental and interstellar components. 
A summary of all features identified is presented in Table~\ref{features}.

A clear signature of the RGS instrumental profile is seen through
absorption structures at 23.05 and 23.35~\AA. These can be interpreted
as the 1s-2p line, broadened and shifted due to the bound state of
oxygen on the instrument. A total column depth of oxygen on the RGS
instrument of $N_{\rm{O}}=2\times10^{17} ~\rm{cm}^{-2}$ is derived.

The RGS instrumental oxygen edge is remarkably stable over time. Early
and later observations of \object{Mrk~421} (orbits 84 and 259) show,
within statistical errors, exactly the same structure and magnitude of
the edge.

The 23.35~\AA~feature is also seen in the Chandra LETG spectra. If the
23.05 and 23.35~\AA~features point to different components, it is clear that
the 23.35~\AA~component is common in both instruments, while the 23.05~\AA~
component is less prominent in Chandra LETG/HRC. 

Likely compounds in the instruments which contain oxygen (apart from
the known oxygen in the Chandra filter polyamide carrier) are
water(ice) and oxidised metal surfaces.

Both XMM-RGS and Chandra LETG have metal surfaces, which will be
slightly oxidised. One could therefore argue that the 23.35~\AA~feature
corresponds to metal oxides.  For RGS, the active Si of the RGS CCD
detectors is covered with a 270~\AA~ $\mathrm{MgF_2}$ insulation layer
which, on top of that, holds an Al light shield, with a thickness in
between 450 and 740~\AA, depending on the CCD (Den Herder et al.
\cite{herder}).  In the Chandra HRC there is a separate filter which holds
an aluminium layer.  In both instruments the oxides can be resident on
the aluminium filters or optical components. For RGS the computed oxygen
column density for the 23.35~\AA~feature corresponds to an aluminium
oxide layer of about 35~\AA. An independent measurement, using
ellipsometry, of the oxide layer on the top of the Al light shield on
one of the flight spare RGS CCDs indicated a layer of only 
$\approx 10$~\AA.
Additional metal oxides will be present on the active Si as well on the
optics.  In the Chandra data the fit of the 23.35~\AA~feature yields an
equivalent aluminium oxide layer thickness of about 110~\AA.

The 23.05~\AA~feature dominates the RGS instrument response.
Contamination (ice forming on the cooled detectors) is very unlikely
the origin of this absorption feature in view of the stable response of
the instrument over a prolonged period of time during repeated
observations of \object{Mrk~421}. In addition, the relative
intensities of the RGS aluminium and fluor calibration sources, which
show X-ray lines at 1.487 and 0.677~keV respectively, did not change
from before launch to recent times (orbit 440). These intensity
ratios are very sensitive to accumulation of oxygen
(water-ice) on the detectors.
  
Another source of oxygen may be absorption of water by the
$\mathrm{MgF_2}$ insulation layer during manufacturing of the CCD's,
before the Al light shield was applied.  This process would lock the
water behind the Al light shield, without any possibility to escape.
$\mathrm{MgF_2}$ is known to be porous, and may be able to absorb
considerable quantities of water, depending on the way it was deposited
on the CCD surface. When all of the oxygen of the 23.05~\AA~feature
would be in the form of water, we would need a layer of 550 \AA~to be
able to account for the observed column density. Despite the fact that
the thickness of the $\mathrm{MgF_2}$ layer is only known for the
``solid equivalent" (assuming that the $\mathrm{MgF_2}$ is not porous),
the required quantity of water seems quite large and is hard
to account for solely in the 270 \AA~$\mathrm{MgF_2}$.

Laboratory measurements of RGS flight-spare CCDs are consistent with
the instrumental oxygen edge measured in flight. The systematic error
for the laboratory measurements, however, is high
compared to the flight measurement. 

The interstellar (ISM) extinction derived from XMM-RGS and Chandra LETG/HRC 
data is fully consistent between the instruments, confirming the validity 
of the analysis.

The interstellar component is compared with theoretical predictions
based on the R-matrix~(McLaughlin and Kirby, \cite{matrix}) computations
of cross sections for atomic oxygen. There is indication that the
theoretical 1s-2p line position requires a slight shift of about
$35\pm10$~m\AA~to larger wavelengths. The structure around the K edge
is shifted relative to the 1s-2p line by $0.13\pm0.02$~\AA~to
longer wavelengths (closer to the 1s-2p line), when compared to the
theory. The overall shape of the edge, which reflects the detailed
resonance structure convolved with the instrument response,
fits the actual data well. The total interstellar X-ray absorption fits the
expected profile for neutral atomic oxygen. No unambiguous additional
signature of oxygen bound in other compounds (e.g. interstellar dust
grains) is found. 
However, we are only sensitive to sufficiently narrow features. If the
type of solid state oxygen in dust grains would exhibit very broad
absorption features we would not recognise it. An upper limit of oxygen
compounds other than neutral oxygen, which have distinct and narrow
features around the oxygen edge is estimated at less than 10~\% of the
neutral oxygen.

A possible interstellar absorption feature of the local intergalactic
medium was found around 22.77~\AA~in both \object{PKS~2155-304} and \object{Mrk~421} spectra.
The exact compound causing this possible absorption feature is not 
unambiguous, but the
fact that this feature is not seen in our galactic sources does point
to an extragalactic origin.
We tentatively identify this feature with \ion{O}{iv}
absorption. 

\begin{acknowledgements}
Based on observations obtained with XMM-Newton and Chandra. XMM-Newton
is an ESA science mission, Chandra is supported by NASA. SRON is supported
financially by NWO, The Netherlands Organisation for Scientific Research.
The Columbia Astrophysics Laboratory is supported by NASA. We like to
thank L. Maraschi for granting permission to use the XMM-RGS data of
orbit 174. The referee is thanked for his useful remarks, which made us 
improve the clarity of the text.  
\end{acknowledgements}

\end{document}